# Multivariate Meta-Analysis: Contributions of Ingram Olkin

**Betsy Jane Becker**

*Abstract.* The research on meta-analysis and particularly multivariate meta-analysis has been greatly influenced by the work of Ingram Olkin. This paper documents Olkin's contributions by way of citation counts and outlines several areas of contribution by Olkin and his academic descendants. An academic family tree is provided.

## 0. INTRODUCTION

Much of the research on statistical methods for meta-analysis in the last three decades has been influenced by Ingram Olkin, either through his direct contributions or through the work of his students and their academic descendants. We indicate the extent of this influence and present a tree of Olkin's academic descendants who have made, or are making, contributions to research in meta-analysis. We then consider the outcome metrics that have been used in the multivariate meta-analysis context and briefly review key results for each metric, thus showing Olkin's seminal influence on this important subfield of meta-analysis.

## 1. OLKIN'S INFLUENCE ON META-ANALYSIS

Meta-analysis is a set of methods for combining and analyzing results from series of related studies. Glass (1976) coined the term "meta-analysis," but the idea of summarizing study results is much older, with references dating to the turn of the last century (e.g., Pearson, 1904). Much of the literature on methods for meta-analysis deals with the univariate case—one endpoint per study. Such endpoints can be represented by correlations, mean differences, proportions, odds ratios (or log odds) and even observed probabilities.

The first of Olkin's contributions to meta-analysis (Hedges and Olkin, 1980) examined the intuitively appealing vote-counting methods used in many research syntheses and traditional literature reviews. Vote counting entails counting the number of studies that have statistically significant results in support of, and counter to, a particular hypothesis, as well as those with nonsignificant results. The category with the most votes (or more than some specific proportion of votes) "wins" and the set of all results is then characterized as supporting that view (e.g., if half of the studies have significant tests in favor of a hypothesis, the studies are viewed as supporting the hypothesis). Hedges and Olkin showed that the statistical properties of this approach were problematic—in that more evidence can lead to poorer decisions.

Since then Olkin has authored or co-authored 39 more articles or book chapters and one book on meta-analysis. The influence of his work is shown by the fact that these documents have generated over 5600 citations. (Based on searches of the Web of Science at http://80isi4.isiknowledge.com.proxy.lib.fsu.edu/ using the author names "Olkin I*," "Hedges L*," "Gleser L*" and "Sampson A*.") His book *Statistical Methods for Meta-analysis* with Larry Hedges (Hedges and Olkin, 1985) is something of a citation classic, having been cited at least 3270 times. However, Olkin's articles and book chapters are also highly cited, with the number of citations per work ranging from 0 to 916 with a mean count of 62.5 citations ($SD = 157.4$) and a median count of 20.5 citations per article. [As is typical of citation counts, the distribution of citation counts per article is highly skewed (skewness coefficient = 4.8), suggesting that the median citation count per article is the more appropriate measure of central tendency.] The majority of this work is collaborative—30 of these papers are co-authored, with the mean number of co-authors across all 39 documents being 2.74 ($SD = 3.5$). As might be expected from the recipient of the Elizabeth L. Scott Award from the Committee of Presidents of Statistical Societies (in 1998), over half (17) of Olkin's 30 co-authored works were written with at least one female co-author.

*Betsy Jane Becker is Professor of Measurement and Statistics, College of Education, Florida State University, Tallahassee, Florida 32306, USA e-mail: bbecker@fsu.edu*

## 2. CONTRIBUTIONS TO META-ANALYSIS OF OLKIN'S ACADEMIC DESCENDANTS

Besides Olkin's own contributions to meta-analysis, individuals that he has mentored and trained have also made many contributions to this literature—some writing dissertations on meta-analysis topics. (Apologies are made to any students of Olkin and his descendants who have inadvertently been omitted from this analysis.) All first-generation descendants were students at Stanford University, though not all earned degrees in the Department of Statistics. In addition, students of those students are considered, and so on, through several generations of Olkin academic "descendants."

These individuals are displayed in Figure 1, the Olkin meta-analytic family tree. The years shown in the figure are the graduation dates for each person; dissertations concerning meta-analysis methods are included in the reference list as well. The tree shows on the bottom-most branches three former students of Olkin who wrote dissertations on meta-analysis. They are Hedges (1980), Holmgren (1989) and Yen (1997). In addition, three other former students of Olkin are shown. Gleser, Perlman and Sampson each have contributed to the literature on meta-analysis or synthesis of results, though none wrote a dissertation on the topic. Relevant works include Gleser and Olkin (1994, 1996), Koziol and Perlman (1978) and Olkin and Sampson (1998), among others.

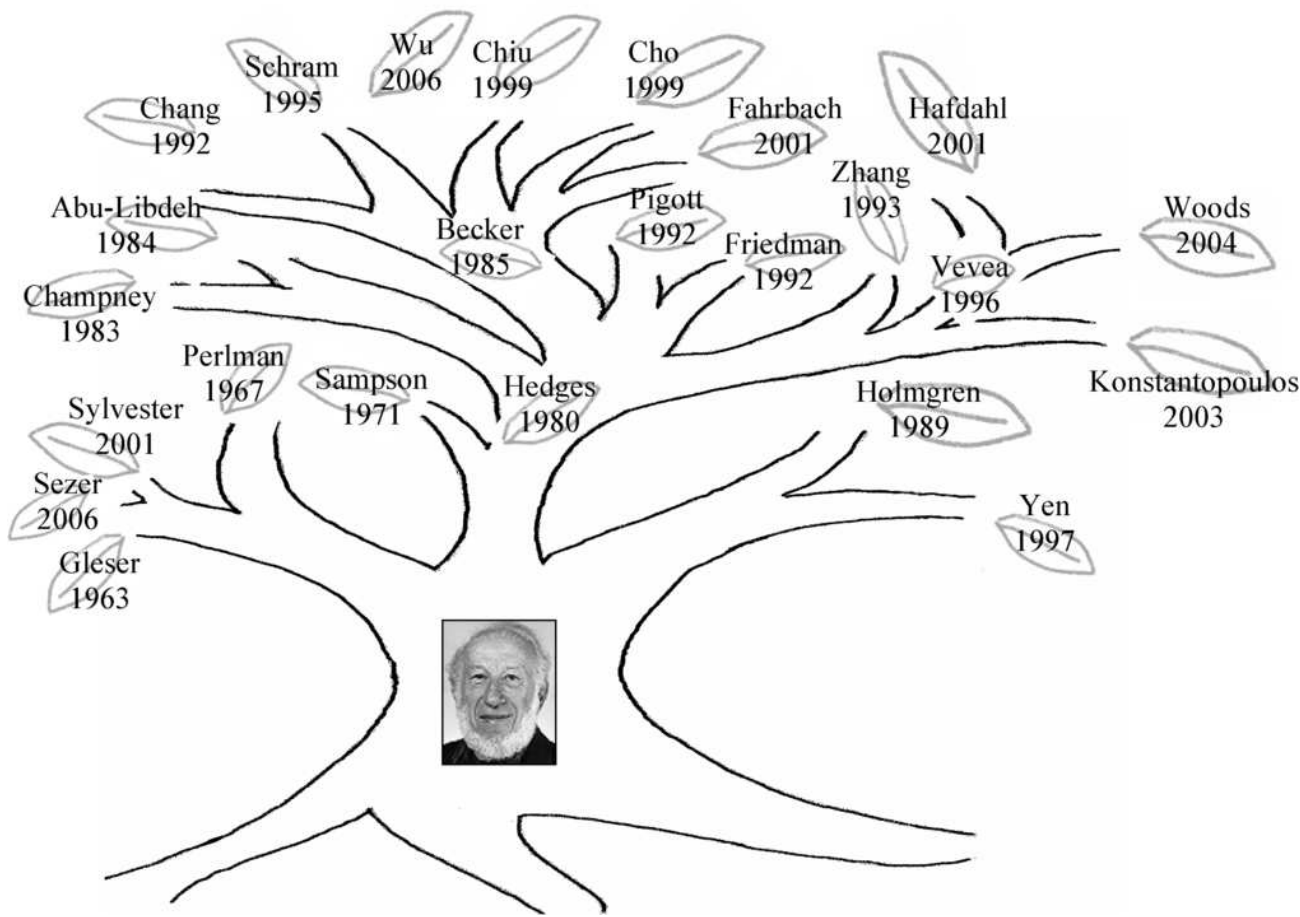

Fig. 1. *The Olkin meta-analytic family tree.*

The next set of leaves shows students of Olkin's students—perhaps we can call these Olkin's meta-analytic "grandchildren." Here are listed seven who wrote dissertations on meta-analytic methods. Abu-Libdeh (1984), Becker (1985), Champney (1983), Konstantopoulos (2003), Pigott (1992) and Zhang (1993) were dissertations written by students of Hedges, and Sylvester (2001) and Sezer (2006) were dissertations directed by Gleser. Two other students of Hedges (Vevea and Friedman) have contributed to the literature on meta-analytic methods after completing a dissertation using meta-analytic methods or on another topic (e.g., Friedman, 1989, 2000; Hedges and Vevea, 1996, 1998).

Finally we reach the current ends of the branches. Six additional students are listed who worked with Becker on meta-analytic methods (Chang, 1992; Chiu, 1999; Cho, 2000; Fahrbach, 2001; Schram, 1996; Wu, 2006) and two who were students of Vevea and who have either written dissertations on meta-analysis methods (Hafdahl, 2001) or contributed to the meta-analytic literature (Vevea and Woods, 2005; Woods et al., 2002) while writing a dissertation on a different topic. We can be assured that others will follow.

## 3. OVERVIEW OF MULTIVARIATE META-ANALYSIS

We next turn to the topic of multivariate meta-analysis and explore Olkin's fundamental contributions to this domain. (See Becker, 2000, and van Houwelingen et al., 2002, for overviews of the topic of multivariate meta-analysis.) Multivariate meta-analysis occurs when more than one (dependent) outcome is measured in a study. This can occur when subjects are measured on several outcomes or at several time points (multiple endpoint studies), or when study indices are computed using shared treatment or control groups (multiple treatment studies). These cases do not typically include studies with results for multiple samples. While such samples may exhibit subtle dependencies because of common instrumentation, treatments and the like, their outcomes do not have a correlation structure that is easily characterized. Hedges and Olkin (1985) first presented methods for dealing with multivariate data in meta-analysis. Their Chapter 10 dealt with standardized mean differences that are dependent because $p$ (dependent) response variables are observed within each primary study.

We denote the results as $T_{ij}$, where $i$ indexes the study and $j$ the outcome. Across studies we may have

$$\begin{bmatrix} T_{11} & \ldots & T_{1p} \\ T_{21} & \ldots & T_{2p} \\ \vdots & & \vdots \\ T_{i1} & \ldots & T_{ip} \\ \vdots & & \vdots \\ T_{k1} & \ldots & T_{kp} \end{bmatrix}$$

for $k$ studies and up to $p$ outcome indices. The $p$ dependent indices arise when $p$ response variables are observed, when contrasts are dependent (e.g., common controls, multiple proportions), when multiple indices involve each response variable (e.g., correlation matrices), and when multivariate analyses appear within a primary study. The possible metrics include multivariate standardized mean differences, correlations and proportions (or odds ratios). Each such metric will be considered in turn.

## 4. MULTIVARIATE STANDARDIZED MEAN DIFFERENCES

This metric may be the most thoroughly investigated of all those for which multivariate analyses have been proposed. Gleser and Olkin (1994) dealt with multiple treatment studies and multiple endpoint studies for standardized mean differences. Some studies combine both of these multivariate aspects. Evidence that multivariate effect-size data are common is found in the fact that Gleser and Olkin (1994) has been cited over 100 times, in fields such as psychology, education, medicine, ecology and criminal justice. Similarly, an early paper by Raudenbush, Becker and Kalaian (1988) dealt with multivariate standardized-mean-difference data.

### 4.1 Multiple Treatment Studies

Multiple treatment studies are illustrated here with an example of studies with a common control group. Further elaborations of this scenario (e.g., with three or more treatment groups or multiple control groups) lead to more outcomes, but the principles underlying these methods can be illustrated with this simplest scenario.

Suppose a study has two treatment groups, $T_1$ and $T_2$, and one control group $C$. Then if we define $\bar{X}^A$ to represent the mean of group $A$ and $S$ to be the pooled within-groups standard deviation across all groups, we can compute

$$T_1 = (\bar{X}^{T_1} - \bar{X}^C)/S \quad \text{and} \quad T_2 = (\bar{X}^{T_2} - \bar{X}^C)/S$$

for each study. If we index these outcomes as $T_{i1}$ and $T_{i2}$ with $i$ for the $i$th study, we will have

$$\begin{bmatrix} T_{11} & T_{12} \\ T_{21} & T_{22} \\ \vdots & \vdots \\ T_{i1} & T_{i2} \\ \vdots & \vdots \\ T_{k1} & T_{k2} \end{bmatrix}$$

which has a multivariate structure. Gleser and Olkin (1994) gave two formulas for $\text{Cov}(T_{ij}, T_{ij'})$ for multiple treatment studies. More recent work by Cook (2004) presents a formula tailored to small-sample cases.

### 4.2 Multiple Endpoint Studies

Gleser and Olkin (1994) also cover dependence of standardized mean differences due to multiple response variables (expanding on Hedges and Olkin, 1985). If we define $T_{ij}$ to represent an effect size for outcome measure $j$ ($j = 1$ to $p$) in study $i$, we have

$$T_{ij} = (\bar{Y}_{ij}^T - \bar{Y}_{ij}^C)/S_{ij}$$

for $i = 1$ to $k$ studies and $j = 1$ to $p$ measures. This was labeled the multiple endpoint design. The effect-size data structure is identical to that shown above but the covariances between the multiple effects from each study differ from those in the multiple treatment case.

## 5. MULTIVARIATE PROPORTIONS

Less has been published on the multivariate meta-analysis of proportions. One contribution is Gleser and Olkin's (2000) chapter on multiple treatment studies with outcomes expressed as two-by-two tables. Gleser and Olkin present large-sample generalized least squares methods for dealing with risk differences, log odds ratios, and arcsine transformed proportions from multiple treatment studies. Other relevant references include Arends, Voko and Stijnen (2003) and Nam, Mengersen and Garthwaite (2003) which concern analyses of multiple log-odds ratios. Additional forthcoming work will undoubtedly address this issue.

## 6. MULTIVARIATE CORRELATIONS AND SLOPES

The topic of synthesis of correlation matrices has seen increasing activity in the past few years. This increase in interest is likely related to the increasingly complex models investigated in primary research, at least in the social sciences. Researchers want to be able to statistically model the effects of multiple predictors as well as to control for potential confounding variables, and this is done by including such variables in complex models. Results of such techniques as structural equation modeling, factor analysis and multiple regression have often been omitted from meta-analyses because of a lack of methods for synthesizing indices from these analyses. While Olkin has not contributed directly to this area of synthesis methods, his work is fundamental because most of the analyses proposed to date are asymptotic and rely on the large-sample distribution theory presented by Olkin and Siotani in 1967.

The multivariate work in this realm of meta-analysis has involved the synthesis of correlation matrices, and the use of those summaries in further modeling of linear models, structural equation models, and even factor analysis (G. Becker, 1996). B. Becker and her collaborators (B. Becker, 1992, 1995; Becker and Fahrbach, 1994; Becker and Schram, 1994) began this stream of work by presenting methods for the synthesis of correlation matrices, specifically estimates of mean matrices under fixed- and random-effects models and tests of the homogeneity of the series of matrices under review. At roughly the same time, applications of like methods appeared in the personnel psychology literature (e.g., Schmidt, Hunter and Outerbridge, 1986). Becker also presented methods for estimating linear models based on the mean correlation matrices and testing components of those composite models. Others have pursued this work and investigated the use of mean matrices with structural equation modeling software (e.g., Cheung and Chan, 2005; Furlow and Beretvas, 2005). All of these works rely on the fundamental result derived by Olkin and Siotani (1976, page 238) of the covariance among correlations from a single sample. Specifically, the large-sample covariance, $\sigma_{ist,iuv}$, between population correlations $\rho_{ist}$ and $\rho_{iuv}$ within study $i$ is

$$\sigma_{r_{ist},r_{iuv}} = [0.5\rho_{ist}\rho_{iuv}(\rho_{isu}^2 + \rho_{isv}^2 + \rho_{itu}^2 + \rho_{itv}^2)$$
$$+ \rho_{isu}\rho_{itv} + \rho_{isv}\rho_{itu}$$
$$- (\rho_{ist}\rho_{isu}\rho_{isv} + \rho_{its}\rho_{itu}\rho_{itv}$$
$$+ \rho_{ius}\rho_{iut}\rho_{iuv} + \rho_{ivs}\rho_{ivt}\rho_{ivu})]/n_i,$$

where $n_i$ is the sample size in study $i$ and $s$, $t$, $u$ and $v$ index the variables within study $i$ that are correlated. That is, $\rho_{ist}$ is the correlation between variables $X_s$ and $X_t$ within study $i$. This result was also used by Hafdahl (2001) who examined exploratory factor analysis methods based on synthesized matrices, and papers by Olkin and other collaborators (e.g., Olkin and Finn, 1976, 1990; Olkin and Saner, 2001) also rely on this fundamental result.

## 7. CONCLUSION

It is safe to say that much of the work on meta-analysis, and especially multivariate issues in meta-analysis, has its genesis in the contributions of Ingram Olkin. The review of research in this paper shows the significant impact of Olkin's work. The family tree illustrates that contributions from Olkin's academic descendants are numerous and will continue to be forthcoming.

## ACKNOWLEDGMENTS

This work was supported by National Science Foundation Grants REC-0335656 and REC-0634013. Parts of this work were presented in the symposium "Multivariate Analysis: In Celebration of Ingram Olkin's 80th Birthday," at the Joint Statistical Meetings, Toronto, Canada, August 2004.